\documentclass[11pt]{article}

\usepackage[preprint]{acl}

\usepackage{times}
\usepackage{latexsym}

\usepackage[T1]{fontenc}
\usepackage[most]{tcolorbox}   
\usepackage{xcolor}
\usepackage[utf8]{inputenc}

\usepackage{microtype}
\usepackage{multicol}
\usepackage{inconsolata}

\usepackage{amsmath}
\usepackage{amssymb}

\usepackage{graphicx}

\usepackage{booktabs}
\usepackage{array}
\usepackage{xcolor}
\usepackage{colortbl}
\usepackage{makecell}
\usepackage{booktabs}   
\usepackage{tabularx}   
\usepackage{array}
\usepackage{makecell}
\usepackage{xcolor}     
\usepackage[dvipsnames]{xcolor} 
\definecolor{promptframe}{RGB}{196, 122, 76}   
\definecolor{promptback}{RGB}{253, 245, 235}   

\newtcolorbox{promptbox}[1]{
    enhanced,
    colback=promptback,           
    colframe=promptframe,         
    colbacktitle=promptframe,     
    coltitle=white,               
    fonttitle=\bfseries,
    title=#1,                     
    sharp corners=south,          
    rounded corners=north,
    boxrule=0.8pt,
    left=8pt, right=8pt, top=6pt, bottom=6pt,
    breakable                     
}

\title{Need to Know: Contextual-Integrity-Grounded Query Rewriting for Privacy-Conscious LLM Delegation}

\author{
  \textbf{Xinyue Huang},
  \textbf{Xiaochun Cao},
  \textbf{Wenyuan Yang}\thanks{Corresponding author.}
  \\[0.5em]
  Sun Yat-sen University \\
  \texttt{huangxy625@mail2.sysu.edu.cn} \quad
  \texttt{\{caoxiaochun,yangwy56\}@mail.sysu.edu.cn}
}

\begin{document}
\maketitle
\renewcommand{\thefootnote}{}
\footnotetext{\noindent Preprint. Work in progress.}
\renewcommand{\thefootnote}{\arabic{footnote}}
\begin{abstract}
As LLMs become increasingly woven into everyday workflows, user queries sent to cloud-hosted LLMs routinely mix\emph{task-essential} content with \emph{task-non-essential} sensitive disclosures, yet type-based PII redaction is context-agnostic and may raise two issues: over-disclosing untyped sensitive context and over-removing answer-bearing spans. We recast privacy-preserving
query rewriting under \emph{Contextual Integrity}: a span should be
forwarded only if it is necessary for the task. We introduce \textbf{DelegateCI-Bench}, the first task-based Contextual Integrity benchmark for privacy-conscious delegation, comprising 3,167 samples that combine high-quality synthetic data spanning 11 tasks and 20 task types, WildChat-based real-user queries, and a medical challenge set with dense sensitive information. Building on this benchmark, we propose a \textbf{CI-guided reinforcement learning framework} that converts essential and non-essential sensitive spans into verifiable optimization signals, and train a query rewriter to preserve task-critical information while suppressing unnecessary sensitive disclosure. Experiments show that our learned rewriter achieves the best privacy--utility trade-off, achieving up to $+10.1$ average utility over on-device baselines.
\end{abstract}
\section{Introduction}
\label{sec:intro}

Users increasingly turn to powerful cloud-hosted LLMs to complete
everyday tasks, but their raw requests rarely contain only
the information that the task requires: alongside the actual
question, users routinely reveal health conditions, family
relationships, financial pressure, or third-party identities.
Forwarding such requests verbatim to an untrusted remote model
creates unnecessary privacy exposure that is wholly extraneous to
what the user asked for.~\citep{neel2023privacy}

\begin{figure}[t]
  \centering
  \includegraphics[width=1\linewidth]{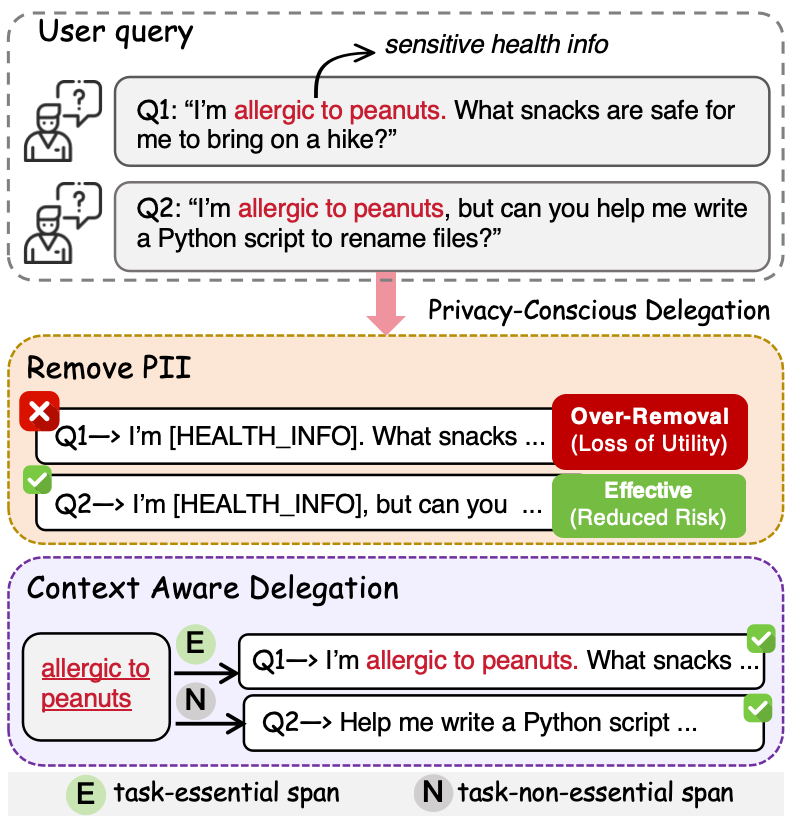}
  \caption{Why type-based PII redaction is insufficient for
  Privacy-Conscious Delegation. }
  \label{fig:teaser}
\end{figure}

We study \textbf{Privacy-Conscious Delegation}, in which a trusted
on-device small model acts as a privacy proxy: it (i)~reads the
user's original request, (ii)~rewrites it into a less-revealing
prompt that still carries the context required to solve the task
and forwards only this prompt to the remote LLM, and (iii)~combines
the original request with the returned remote answer to produce the
final response~\citep{papillon}. This goes beyond standard PII
redaction~\citep{prediax,pilan2022text}: the local
model performs task-aware context selection and rewriting
rather than scrubbing identifiers by type.

The central difficulty of this rewriting step is task
dependence: whether a span should be disclosed is determined not
by whether it is sensitive in isolation, but by whether it is useful
for the current task. As shown in Figure~\ref{fig:teaser}, ``\emph{I'm
allergic to peanuts}'' must be preserved when asking for
hike-friendly snacks, yet should be removed when asking for a
Python file-renaming script. Type-based PII redaction, which strips
spans by category regardless of context, cannot make this
distinction and systematically errs in both directions.

We therefore adopt \textbf{Contextual Integrity}
(CI)~\citep{nissenbaum2010privacy} as our guiding principle. CI
holds that privacy is not determined by information type alone, but
by whether information flows appropriately given the actors,
context, and purpose involved. We operationalize CI for our setting
as task necessity for remote assistance: a span should be
disclosed to the remote model if and only if that model needs it to
provide effective task support.

To make this criterion directly measurable, we build a CI-grounded
benchmark in which every request is annotated with two complementary
span sets: \textbf{task-essential spans} $\mathcal{E}$ and
\textbf{task-non-essential sensitive spans} $\mathcal{N}$. These two
sets pin down two dual failure modes: forwarding any span in
$\mathcal{N}$ is \textbf{over-disclosure}, while stripping any span
in $\mathcal{E}$ is \textbf{over-removal}. The benchmark draws on
three complementary sources. The general-domain split is built by
reverse synthesis---a sampler-generator-judge closed loop
that fixes the $(\mathcal{E},\mathcal{N})$ partition before
rendering the surface query, yielding labels correct by construction.
We further anchor the distribution with a natural slice from
WildChat~\citep{zhao2024wildchat} and add a medical challenge set
that stresses over-removal.

We further turn $(\mathcal{E},\mathcal{N})$ into a verifiable
RL reward and train the delegation agent with Group Relative Policy
Optimization (GRPO)~\citep{shao2024deepseekmath}. The composite
reward treats privacy as a \textbf{hard constraint}---any disclosure
of a span in $\mathcal{N}$ saturates the reward to its lower
bound---rewards retention of $\mathcal{E}$ via a smoothed recall
term, and adds a brevity penalty against reward hacking. A single
rewriter checkpoint is shared across heterogeneous downstream
aggregators, so the learned behaviour is a
backbone-agnostic, context-dependent disclosure policy. We summarize our contributions as follows:
\begin{itemize}
    \item We recast Privacy-Conscious Delegation through the lens of
    Contextual Integrity, shifting the operative criterion from
    type-based ``is this span PII?'' to task-conditioned ``does this
    span carry task-solving context for the remote LLM?''.
    \item We construct \textbf{DelegateCI-Bench}, the first task-based CI benchmark for Privacy-Conscious Delegation, spanning broad controllable general-domain queries, naturalistic privacy leakage in real user prompts, and challenge cases where sensitive information is itself task-bearing evidence, thereby testing whether a rewriter can preserve task-critical context while removing or generalizing unnecessary sensitive disclosures.
    \item We show that the span partition in \textbf{DelegateCI-Bench} can directly supervise a privacy-aware query rewriter. This converts contextual integrity from an evaluation principle into a trainable objective, yielding a single GRPO-trained rewriter with the strongest privacy--utility trade-off.
\end{itemize}

\section{Related Work}
\label{sec:related}

\subsection{Privacy Protection for User Prompts}
\label{sec:related:prompt-privacy}
One line of work mitigates prompt-side leakage via
input anonymization or perturbation, sanitizing the request
locally and optionally restoring redacted entities after the cloud
call~\citep{kan2023ppts,chen2023hideandseek}; such pipelines act on
predefined sensitive categories and cannot judge whether a
span is sensitive given the current task. A parallel
differential-privacy thread provides formal guarantees but is largely
confined to ICL with reusable demonstrations and does not extend to
free-form requests~\citep{tang2024privacy,hong2024dpopt}.
Closest to our setting, \citet{papillon} formulate
Privacy-Conscious Delegation via a trusted local proxy, but
reduce privacy to PII leakage. We retain the delegation paradigm
but reframe disclosure through Contextual Integrity,
supervising the rewriter with a verifiable
$(\mathcal{E},\mathcal{N})$ partition (Sec.~\ref{sec:method}) and
jointly evaluating over-disclosure of non-essential spans and
over-removal of task-essential ones.

\subsection{Contextual Integrity for LLM Privacy}
\label{sec:related:ci}
Existing CI-grounded benchmarks~\citep{nissenbaum2010privacy}
primarily diagnose the output side, asking whether a remote
model relays information to inappropriate
recipients~\citep{mireshghallah2024can,shao2024privacylens,cheng2024cibench}.
\citet{ngong2025contextual} shift attention to the input side
and propose inference-time local rewriting. We adopt this stance
but operationalise CI as a span-level $(\mathcal{E},\mathcal{N})$
partition that doubles as a verifiable training reward and a
benchmark axis, enabling end-to-end learned disclosure and the first
joint evaluation of over-disclosure and over-removal.
\section{DelegateCI-Bench Construction}
\label{sec:data}

Existing privacy benchmarks for LLMs are largely PII-centric and
treat sensitivity as a property of information
types~\citep{pilan2022text,lukas2023analyzing}. This type-based
view is insufficient for Privacy-Conscious Delegation, where the privacy decision is not simply whether a piece of information is sensitive, but whether it is necessary to
disclose for the delegated task. Building on the Contextual Integrity
(CI) framing introduced there, we focus on task-conditioned information
disclosure: whether a sensitive attribute should flow to the remote LLM
under the task's transmission principle. We operationalize this idea as
a concrete sample-level annotation protocol and construct DelegateCI-Bench, the first task-based CI benchmark for Privacy-Conscious
Delegation.
\paragraph{Contextual integrity for delegation.}
A CI disclosure decision is governed by five parameters~\citep{nissenbaum2010privacy}: the sender, recipient, subject, information type, and transmission principle. In Privacy-Conscious Delegation, these correspond respectively to the user, the remote LLM, the person whom the information is about, the sensitive attribute being disclosed, and task necessity. Concretely, every sample is annotated with two disjoint
attribute sets: \textbf{essential} spans
$\mathcal{E}=\{e_i\}_{i=1}^{n}$, which are necessary for the task and
should be forwarded, and \textbf{non-essential} spans
$\mathcal{N}=\{c_j\}_{j=1}^{K}$, which are sensitive but not
task-relevant and should be removed or generalized.

\paragraph{Complementary benchmark design.}
To cover both CI failure modes (over-disclosure and over-removal) with
reliable labels and realistic distributions, we assemble (i) a
\textbf{reverse-synthesised general-domain} subset
(\S\ref{sec:data:reverse}) for large-scale, controllable coverage with
labels correct by construction; (ii) a \textbf{WildChat-anchored} OOD
slice of organic single-turn prompts (\S\ref{sec:data:reverse}); and
(iii) a \textbf{medical challenge set} (\S\ref{sec:data:medical}) that stress-tests over-removal where sensitive clinical spans are themselves the answer-bearing evidence. Detailed prompts and value banks are deferred to the appendix. Table~\ref{tab:data-stats} summarises the benchmark, comprising \textbf{3{,}167} samples across two groups---\textbf{General} (reverse-synthesised samples plus the WildChat-anchored OOD probe; \S\ref{sec:data:reverse}) and \textbf{Medical}.

\subsection{General-Domain Samples}
\label{sec:data:reverse}

\paragraph{Why reverse synthesis.}
Prior work~\citep{ngong2025contextual} identified only 2{,}849 queries containing task-irrelevant sensitive information from 11{,}305 filtered single-turn ShareGPT~\citep{sharegpt2023} conversations. This sparsity becomes more pronounced when further requiring coverage across diverse domains, task types, sensitive attributes, information subjects, and contrastive necessity variations of the same attribute across different tasks. We therefore adopt data synthesis to construct multi-domain, multi-task, multi-attribute, and contrastive-necessity samples in a controllable manner. 

To prevent the synthesized data from deviating from real user distributions, we additionally curate a real-world validation subset from WildChat~\citep{zhao2024wildchat}: starting with $\sim$50K single-turn English queries filtered from 352K real conversations, a GPT-4o-mini scorer selects 656 queries exhibiting task-irrelevant privacy exposure, which are then annotated by GPT-4o with $(\mathcal{E}, \mathcal{N})$, yielding 345 final samples. Rather than exhaustively covering all parameter combinations, this real subset serves as an independent authenticity and robustness validation set.

\paragraph{Structured sampler.}
The sampler jointly varies five axes that determine when a sensitive
disclosure is contextually appropriate (Table~\ref{tab:coverage}); the
taxonomy follows \citet{ngong2025contextual}.

\begin{table}[h]
  \centering
  \footnotesize
  \setlength{\tabcolsep}{4pt}
  \renewcommand{\arraystretch}{1.1}
  \caption{Coverage axes of the structured sampler. Each generated
  sample is a joint draw over these axes; the CI judgement is then
  rendered into natural language and re-checked by an independent
  judge (\S\ref{sec:data:reverse}).}
  \label{tab:coverage}
  \begin{tabular}{l c p{0.55\linewidth}}
    \toprule
    \textbf{Axis} & \textbf{\#} & \textbf{Examples} \\
    \midrule
    Domain            & 11 & health, finance, employment, education, \ldots \\
    Task type         & 20 & summarisation, translation, code generation, \ldots \\
    Sensitive attribute & 30 & family history, medication, salary, religion, \ldots \\
    Subject           & 5  & self, family, friend, colleague, third party \\
    Disclosure style  & 4  & direct, narrative, implicit, multi-sentence dispersed \\
    \bottomrule
  \end{tabular}
\end{table}

\paragraph{Generator and consistency judge.}
A generator LLM is conditioned on the sampled CI skeleton and instructed
to render the instance as a natural prompt while embedding the
non-essential attributes. To reduce co-adaptation, we use heterogeneous
models, with Claude Sonnet~4.6 as the generator and GPT-4o as the
consistency judge.
The judge observes only the surface prompt and re-extracts
$(\hat{\mathcal{E}},\hat{\mathcal{N}})$; only samples whose re-extracted
partition agrees with the sampled partition are retained. Detailed
judging prompts, matching rules, thresholds, and human agreement results
are reported in Appendix~\ref{app:judge}.


\subsection{Medical Challenge Set}\label{sec:data:medical}

To keep medical facts grounded, we build the over-removal stress test
on board-exam-style multiple-choice items drawn from the
\textsc{medical-o1-reasoning-SFT} corpus released with
HuatuoGPT-o1~\citep{chen2024huatuogpt_o1}, rather than synthesising
clinical stems from scratch.

For each item we use its existing chain-of-thought solution and
verified answer to drive GPT-4o to \textbf{verbatim} extract from the
original question stem the spans $\mathcal{E}$ that are decisive for
the final diagnosis or option choice. We then use GPT-4o to rewrite each item under a strictly constrained instruction that appends or embeds synthetic, clinically inert disclosures forming $\mathcal{N}$.


\begin{table}[t]
  \centering
  \footnotesize
  \setlength{\tabcolsep}{4pt}
  \renewcommand{\arraystretch}{1.05}
  \caption{DelegateCI-Bench statistics.}
  \label{tab:data-stats}
  \begin{tabular}{l c c}
    \toprule
     & \textbf{General} & \textbf{Medical} \\
    \midrule
    \# samples                                  & 2{,}217 & 950 \\
    Avg.\ $|\mathcal{E}|$ (spans)               & 3.94 & 3.83 \\
    Avg.\ $|\mathcal{N}|$ (spans)               & 4.78 & 3.34 \\
    Avg.\ query length (words)                  & 98.3 & 89.3 \\
    \bottomrule
  \end{tabular}
\end{table}

\section{Method}
\label{sec:method}

In this section, we formalize the privacy-preserving query rewriting
task as a reinforcement learning (RL) problem
(\S\ref{sec:framework}), present a composite reward that jointly
enforces non-essential privacy non-disclosure and task utility
(\S\ref{sec:reward}), and describe the on-policy training procedure
based on Group Relative Policy Optimization (GRPO)
\citep{shao2024deepseekmath}(\S\ref{sec:policy}).

\subsection{Framework Overview}
\label{sec:framework}
\begin{figure*}[t]
  \centering
  \includegraphics[width=1\linewidth]{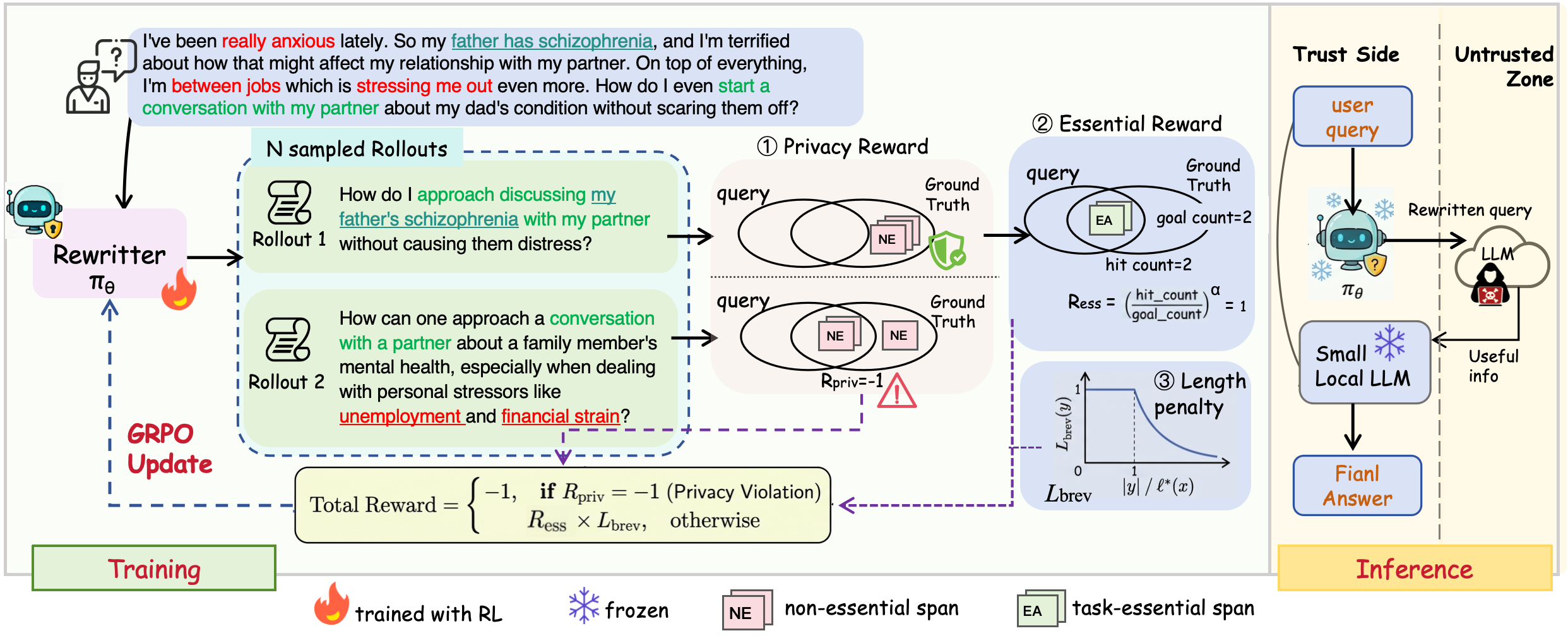}
  \caption{Overview of our framework. \textbf{Training (left):}
  rollouts from $\pi_\theta$ are scored by a composite reward
  combining a hard privacy term $R_{\text{priv}}$, a concave essential
  term $R_{\text{ess}}$, and a one-sided length penalty
  $L_{\text{brev}}$, which drives a GRPO update of $\pi_\theta$.
  \textbf{Inference (right):} the frozen $\pi_\theta$ rewrites the
  query on the trusted side; only the rewrite is sent to the untrusted
  remote LLM, whose response is recombined locally to produce the
  final answer.}
  \label{fig:framework}
\end{figure*}
\paragraph{Problem setup.}
Given a user query $x$, which may contain both information necessary
to complete the task and sensitive information unrelated to the task,
the policy model $\pi_\theta$ generates a rewritten query
$y \sim \pi_\theta(\cdot \mid x)$. This rewritten query is forwarded
to an external untrusted retriever or a remote LLM, as shown in Figure~\ref{fig:framework}. The optimization
objective is to retain as much of the information required to
complete the original task as possible while minimizing the leakage
of unnecessary sensitive information.

We frame this problem as a single-step \emph{contextual bandit}
\citep{langford2007epoch}: each
rollout produces a complete rewrite $y$ that is evaluated by a
sequence-level reward $R(y)$, with no intermediate states or
bootstrapped value estimates. Although $y$ is generated
autoregressively at the token level, the reward in
\S\ref{sec:reward} is intrinsically a holistic property of the full
rewrite (a single leaked span suffices to violate Contextual
Integrity), so we assign the same group-relative advantage uniformly
to every token of $y$ during policy optimization.

\paragraph{Pipeline.}
At inference time, the rewritten query $y$ is sent to an untrustworthy
external large language model; the retrieved evidence is then
recombined with the original $x$ on the trusted side to produce the
final answer. Our scope is limited to learning $\pi_\theta$ that
produces $y$; the downstream answerer is frozen and is not part of
the optimization loop.

\subsection{Reward Design}
\label{sec:reward}

The reward has three components targeting the two Contextual
Integrity failure modes from \S\ref{sec:intro} and one RL-specific
failure: $R_{\text{priv}}$ for over-disclosure, $R_{\text{ess}}$ for
over-removal, and a multiplicative brevity factor $L_{\text{brev}}$
against reward hacking \citep{skalse2022reward} by padding the rewrite with extraneous
content to opportunistically cover $\mathcal{E}$.

\paragraph{Notation.}
Given an input query $x$ and a rewritten query $y$, let
$\mathcal{E}=\{e_i\}_{i=1}^{n}$ be the set of \emph{essential} spans
and $\mathcal{N}=\{c_j\}_{j=1}^{K}$ the set of \emph{non-essential}
spans. For two strings $a, b$, define the token-level Jaccard coverage
\begin{equation}
  \mathrm{cov}(a, b) \;=\; \frac{|T(a) \cap T(b)|}{|T(b)|},
\end{equation}
where $T(\cdot)$ is the lowercased, stop-word-filtered, alphanumeric
token multiset. Let $P_{\text{r}}(a, b) \in [-1, 1]$ denote the
rescaled BERTScore precision \citep{zhang2020bertscore} computed with
a \texttt{roberta-large} backbone.

\paragraph{Privacy reward $R_{\text{priv}}$.}
Relying on a single match criterion leaves the model vulnerable to evasion via paraphrasing or verbatim insertion of short tokens. Thus, we penalize leakage using a disjunction of strict lexical coverage and semantic similarity. To prevent the over-removal of spans that are both sensitive and answer-bearing (\S\ref{sec:intro}), we first remove any $c \in \mathcal{N}$ that lexically overlaps with spans in $\mathcal{E}$, yielding a filtered set $\mathcal{N}^\prime \subseteq \mathcal{N}$. The reward is formulated as:
\begin{equation}
\resizebox{0.89\linewidth}{!}{$
  R_{\text{priv}}(y) =
  \begin{cases}
    -1, & \exists\, c \in \mathcal{N}^\prime:\\
        & \mathrm{cov}(y, c) \geq 1.0 \lor P_{\text{r}}(y, c) > \tau_{\text{priv}},\\[2pt]
    \phantom{-}0, & \text{otherwise}.
  \end{cases}
$}
\end{equation}
We use a strict lexical threshold of $1.0$ to avoid penalizing benign partial overlaps (e.g., common function words). The semantic threshold $\tau_{\text{priv}}=0.3$ is empirically calibrated on a held-out development set. Finally, an empty sequence $y$ is penalized with $R_{\text{priv}}=-1$ to discourage trivial silent behaviors.

\paragraph{Essential-preservation reward.}
Conditional on $R_{\text{priv}}(y)=0$, we measure utility by the
fraction of $\mathcal{E}$ retained by $y$:
\begin{equation}
  \rho(y) \;=\; \frac{1}{|\mathcal{E}|}
              \sum_{e \in \mathcal{E}} \phi(y, e),
\end{equation}
where $\phi(y,e)=\mathbf{1}[\mathrm{cov}(y,e)\!\geq\!1.0\,\lor\,
P_{\text{r}}(y,e)\!>\!\tau_{\text{priv}}]$. With a linear reward
$\rho$, recovering one additional essential span is worth a constant
$1/|\mathcal{E}|$, which can become vanishingly small when
$|\mathcal{E}|$ is large. As a result, early-stage training may
receive only a weak learning signal. We therefore reshape $\rho$ with a \emph{concave} exponent:
\begin{equation}
  R_{\text{ess}}(y) \;=\; \rho(y)^{\,\alpha},\qquad \alpha\in(0,1),
\end{equation}
which by construction (a) is monotonically increasing in $\rho$, so
the global optimum still lies at full recall; and (b) has strictly
decreasing marginal returns,
$\partial R_{\text{ess}}/\partial \rho = \alpha\,\rho^{\alpha-1}$,
which diverges as $\rho\to 0^{+}$ and decays to $\alpha$ at
$\rho=1$. This amplifies credit for the first few recovered spans and
densifies the early-training signal.


\paragraph{Length penalty.}
A policy can pad a leak-free rewrite with task-irrelevant but privacy-safe
material to maximize $\rho(y)$. We block this
with a one-sided, multiplicative brevity factor anchored at a
per-example target length $\ell^\star(x) = |y^\star(x)|$, where
$y^\star(x)$ is the target rewrite pre-annotated by an advanced model during
benchmark construction and used as the desired upper bound on length.
This upper bound adapts to query length and information density.
The brevity factor is
\begin{equation}
  L_{\text{brev}}(y) \;=\;
  \min\!\left(1,\;
    \exp\!\Bigl(1 - \tfrac{|y|}{\ell^\star(x)}\Bigr)
  \right)
  \;\in\;(0, 1].
\end{equation}
The penalty is intentionally one-sided: overly long rewrites are penalized.

\paragraph{Final reward.}
Combining the three terms,
\begin{equation}
  R(y) =
  \begin{cases}
    R_{\mathrm{ess}}(y)\,L_{\mathrm{brev}}(y), & R_{\mathrm{priv}}(y)=0,\\[2pt]
    -1, & R_{\mathrm{priv}}(y)=-1.
  \end{cases}
\end{equation}
The multiplicative form ensures brevity can only \emph{discount}
retained utility, and any leak forces $R(y){=}{-}1$ regardless of
recall or length.

\subsection{Policy Training}
\label{sec:policy}

We optimize $\pi_\theta$ with GRPO \citep{shao2024deepseekmath}, which adopts a PPO-style clipped surrogate objective \citep{schulman2017ppo}, using reward $R(y)$. For each
query $x$ we draw $G$ rollouts $\{y_i\}_{i=1}^{G}$ from $\pi_\theta$
and compute their rewards $\{R_i\}_{i=1}^{G}$. The group-relative
advantage is
\begin{equation}
\label{eq:adv}
  \hat{A}_i \;=\;
  \frac{R_i - \mathrm{mean}(\{R_j\}_{j=1}^{G})}
       {\mathrm{std}(\{R_j\}_{j=1}^{G}) + \epsilon}.
\end{equation}
Let
$r_i(\theta) = \pi_\theta(y_i \mid x) / \pi_{\theta_{\text{old}}}(y_i \mid x)$, the GRPO objective is
\begin{equation}
\small
\begin{aligned}
\mathcal{L}_{\text{GRPO}} = -\mathbb{E}\Bigl[&\tfrac{1}{G}\!\sum_{i=1}^{G}\min\!\bigl(r_i\hat{A}_i,\mathrm{clip}(r_i,1{\pm}\epsilon_c)\hat{A}_i\bigr) \\
&- \beta D_{\mathrm{KL}}(\pi_\theta\|\pi_{\text{ref}})\Bigr].
\end{aligned}
\end{equation}
where $\epsilon_c$ is the clipping range and $\beta$ the KL
coefficient. Per the contextual-bandit framing, $\log \pi_\theta(y_i \mid x)$ factorizes
over tokens and the same $\hat{A}_i$ applies to each token of $y_i$.
\section{Experiments}
\label{sec:experiments}
\newcommand{\best}[1]{\textbf{#1}}                  
\newcommand{\sbest}[1]{\underline{#1}}              
\newcommand{\aggrhead}[2]{%
  \rowcolor{#1}\multicolumn{12}{c}{\textbf{#2}}\\}
\newcommand{\todofill}{{\color{gray!70}?}}          
\definecolor{ourrow}{HTML}{FFF6D6}                  
\definecolor{qwenblock}{HTML}{FFF1E6}               
\definecolor{llamablock}{HTML}{EAF7EA}              
\definecolor{metricgray}{HTML}{F5F5F5}              
\definecolor{refblock}{HTML}{EEF2FB}                

\begin{table*}[t]
\centering
\setlength{\tabcolsep}{6pt}
\renewcommand{\arraystretch}{1.22}
\caption{%
  \textbf{Privacy--utility comparison of query rewriting strategies under three
  frozen local aggregators.} \best{Bold} and \sbest{underline} mark the best
  and second-best \emph{protected} results within each aggregator block.
} 
\label{tab:main}
\small
\resizebox{\textwidth}{!}{%
\begin{tabular}{l l c c c c c c c c c c}
\toprule
\textbf{Methods} & \textbf{Rewriter}
& \multicolumn{4}{c}{\textbf{General}}
& \multicolumn{4}{c}{\textbf{Medical}}
& \textbf{Avg.\,Util.}\,$\uparrow$
& \textbf{$\Delta$ vs.\,Local}\,$\uparrow$ \\
\cmidrule(lr){3-6}\cmidrule(lr){7-10}
& &
SLR\,$\downarrow$ &
NE\,$\downarrow$ &
EA\,$\uparrow$ &
Util.\,$\uparrow$ &
SLR\,$\downarrow$ &
NE\,$\downarrow$ &
EA\,$\uparrow$ &
Util.\,$\uparrow$ & & \\
\midrule

Direct LLM & raw query
& \multicolumn{3}{c}{\textit{unprotected}} & 100.0$^{\star}$
& \multicolumn{3}{c}{\textit{unprotected}} & 79.5
& 89.8 & --- \\
Presidio(PII Only) & rule-based PII
& 89.6 & 53.7 & 96.2 & 89.2
& 1.8 & 0.6 & 95.3 & 64.2
& 76.7 & --- \\

\midrule
\aggrhead{qwenblock}{Local aggregator: \textit{Qwen-2.5-1.5B-Instruct}}
Local Only & original query
& \multicolumn{3}{c}{\textit{on-device}} & 71.8
& \multicolumn{3}{c}{\textit{on-device}} & 34.0
& 52.9 & 0.0 \\
Papillon & Qwen-2.5-1.5B
& 38.5 & 25.5 & \best{61.5} & \sbest{80.3}
& 62.7 & 40.8 & \sbest{61.5} & \sbest{35.0}
& \sbest{57.7} & \sbest{$+$4.8} \\
PUFT & Qwen-2.5-1.5B
& \sbest{19.8} & \sbest{12.0} & 44.5 & 80.0
& \sbest{16.7} & \sbest{9.9} & 13.1 & 32.5
& 56.3 & $+$3.4 \\
\rowcolor{ourrow}
\textbf{Ours} & Qwen-2.5-3B (\emph{shared})
& \best{8.0} & \best{2.2} & \sbest{59.8} & \best{89.5}
& \best{4.2} & \best{1.5} & \best{71.0} & \best{41.2}
& \best{65.4} & \best{$+$12.5} \\

\midrule
\aggrhead{qwenblock}{Local aggregator: \textit{Qwen-2.5-7B-Instruct}}
Local Only & original query
& \multicolumn{3}{c}{\textit{on-device}} & 89.9
& \multicolumn{3}{c}{\textit{on-device}} & 47.3
& 68.6 & 0.0 \\
Papillon & Qwen-2.5-7B
& \best{6.9} & \best{2.0} & \best{63.4} & 93.5
& 55.0 & 23.9 & \best{80.6} & \sbest{59.0}
& \sbest{76.3} & \sbest{$+$7.7} \\
PUFT & Qwen-2.5-7B
& \sbest{7.6} & 2.5 & 50.1 & \best{96.0}
& \sbest{20.3} & \sbest{7.2} & 27.8 & 49.0
& 72.5 & $+$3.9 \\
\rowcolor{ourrow}
\textbf{Ours} & Qwen-2.5-3B (\emph{shared})
& 8.0 & \sbest{2.2} & \sbest{59.8} & \sbest{95.3}
& \best{4.2} & \best{1.5} & \sbest{71.0} & \best{62.0}
& \best{78.7} & \best{$+$10.1} \\

\midrule
\aggrhead{llamablock}{Local aggregator: \textit{Llama-3.1-8B-Instruct}}
Local Only & original query
& \multicolumn{3}{c}{\textit{on-device}} & 87.2
& \multicolumn{3}{c}{\textit{on-device}} & 60.7
& 74.0 & 0.0 \\
Papillon & Llama-3.1-8B
& \best{3.0} & 4.6 & \sbest{52.4} & 92.2
& 46.7 & 16.7 & \sbest{61.7} & 49.2
& 70.7 & $-$3.3 \\
PUFT & Llama-3.1-8B
& \sbest{5.1} & \sbest{2.4} & 40.6 & \sbest{92.7}
& \sbest{28.5} & \sbest{9.5} & 41.9 & \sbest{62.8}
& \sbest{77.8} & \sbest{$+$3.8} \\
\rowcolor{ourrow}
\textbf{Ours} & Qwen-2.5-3B (\emph{shared})
& 8.0 & \best{2.2} & \best{59.8} & \best{94.3}
& \best{4.2} & \best{1.5} & \best{71.0} & \best{66.7}
& \best{80.5} & \best{$+$6.6} \\
\bottomrule
\end{tabular}%
}

\vspace{2pt}
{\scriptsize
$^{\star}$\,Utility ceiling, not a protected method.\quad
``\textit{on-device}'' indicates that no information is exposed off-device,
so privacy metrics are not applicable.
}
\end{table*}
We evaluate the proposed privacy-preserving query rewriter on two
complementary axes. \emph{Privacy} measures how much
task-non-essential sensitive content survives in the rewritten query
that is exposed to the remote LLM, while \emph{utility} measures
whether the rewrite is still informative enough for a frozen local
aggregator to produce a high-quality response.

\subsection{Experimental Setup}
\label{sec:exp-setup}

\paragraph{Models.}
We instantiate the policy rewriter $\pi_\theta$ from
\textsc{Qwen-2.5-3B-Instruct} and train it with GRPO under the reward
of \S\ref{sec:reward}. A \emph{single} rewriter checkpoint is shared
across all evaluation settings, testing whether the learned rewriting
behaviour generalises beyond a particular backbone. The remote
answerer is \textsc{GPT-4o}, accessed through an internal proxy. On
the trusted side, we evaluate three frozen local aggregators of
increasing scale: \textsc{Qwen-2.5-1.5B-Instruct},
\textsc{Qwen-2.5-7B-Instruct}, and \textsc{Llama-3.1-8B-Instruct}.
Training details are reported in Appendix~\ref{app:training}.

\paragraph{Evaluation data.}
We evaluate on DelegateCI-Bench \S\ref{sec:data}, in which every test
instance carries an explicit $(\mathcal{E},\mathcal{N})$ partition of
essential and non-essential sensitive spans. For evaluation, we report results on 1600 held-out portions of two complementary groups: \textbf{General} and \textbf{Medical}.

\paragraph{Evaluation metrics.}

\textit{Privacy.} An item $c\in\mathcal{N}$ is flagged as leaked in a
rewrite $y$ if it is recoverable from $y$ via either token-level
Jaccard coverage or rescaled BERTScore
precision~\citep{zhang2020bertscore}, capturing
both verbatim and paraphrastic disclosure; items already in
$\mathcal{E}$ are excluded. We report
\textbf{SLR} (sample-level non-essential leak rate, our headline
privacy score), \textbf{NE} (per-item leak rate over $\mathcal{N}$),
and \textbf{EA} (per-item retention rate over $\mathcal{E}$). Lower
SLR/NE and higher EA are better.

\textit{Utility.} On \emph{General} we use the pairwise-with-swap
judge of \citet{papillon}: the local-pipeline answer $C_L$ (driven by
the rewrite $y$) is compared in both orderings against the unprotected
\textsc{GPT-4o} ceiling answer $C_T$, and \textit{Util.} is the
swap-consistent win-or-tie rate. On \emph{Medical}, where each item
has a verified gold option, \textit{Util.} is exact-match accuracy.

\paragraph{Baselines.}
\textbf{Direct LLM} forwards the unprotected query to \textsc{GPT-4o}
and serves as the utility ceiling (not comparable on the privacy
axis). \textbf{Local Only} runs the original query entirely on the
frozen local aggregator and serves as utility floor. We further compare against two query-side defences:
\textbf{Presidio}~\citep{prediax}, a rule-based PII redactor, and
\textbf{PAPILLON}~\citep{papillon}, a PII-oriented privacy-conscious delegation baseline; and \textbf{PUFT}~\citep{ngong2025contextual}, a conversational-agent-oriented local reformulation method that removes or rewrites non-essential sensitive information while preserving user intent. Details are in Appendix~\ref{app:baselines}.

\subsection{Main Results on Privacy-Preserving Query Rewriting}
\label{sec:exp-main}
Table~\ref{tab:main} reports the privacy--utility trade-off across
three aggregator scales. Our rewriter achieves the best average utility
among protected methods in every block, peaking at $80.5$ with
\textsc{Llama-3.1-8B} and yielding the largest gain over Local Only
($+12.5$) in the most constrained \textsc{Qwen-2.5-1.5B} setting. Because the same rewriter checkpoint is reused across all aggregators, these gains are not tied to a specific downstream backbone. 

The improvement is most pronounced on the \emph{Medical} split. With \textsc{Qwen-2.5-7B}, our method lifts Medical utility from $47.3$ (Local Only) to $62.0$, while reducing leakage relative to PAPILLON; with \textsc{Llama-3.1-8B}, it reaches $66.7$, versus $60.7$ for Local
Only and $62.8$ for PUFT. This pattern reflects the structure of the
Medical split: answer-decisive information is often concentrated in
essential sensitive spans. Since our rewriter is trained to preserve
$\mathcal{E}$ while removing $\mathcal{N}$, it can retain the clinical
context needed for the correct answer. PAPILLON can reduce surface leakage in larger-model settings, but its aggressive paraphrasing makes performance less consistent across aggregator scales. PUFT is more privacy-conservative, yet often removes task-relevant evidence, particularly on Medical where essential information overlaps with sensitive spans. Overall, these results highlight the need to jointly preserve essential information and suppress non-essential
private content.
\begin{table}[t]
  \centering
  \footnotesize
  \setlength{\tabcolsep}{3pt}
  \caption{Privacy-leak rates (\%) by category. 
Direct, semantic, inferential, and combination leaks denote verbatim retention, recoverable paraphrase, contextual inference, and re-identification from combined cues, respectively. 
$\ge\!1$\,/\, $\ge\!2$ denote rewrites with at least one\,/\,two leak categories.}
  \label{tab:leaks}
  \begin{tabular}{lcccccc}
    \toprule
    System   & direct       & semantic     & infer.       & comb.         & $\ge 1$         & $\ge 2$        \\
    \midrule
    Papillon & \textbf{15.8} & 4.2          & 12.8         & 50.8          & 76.4            & 7.2            \\
    Prediax  & 51.8          & 2.6          & 10.6         & \textbf{34.6} & 86.2            & 12.2           \\
    PUFT     & 21.0          & 2.4          & \textbf{7.5} & 39.2          & 70.4            & 6.9            \\
    \textbf{Ours} & 23.0     & \textbf{2.2} & 8.0          & 39.6          & \textbf{68.4}   & \textbf{4.2}   \\
    \bottomrule
  \end{tabular}
\end{table}
\subsection{LLM-as-a-Judge Assessment}
\label{sec:llm-judge}
\begin{table*}[t]
    \centering
    \small
    \renewcommand{\arraystretch}{1.25}
    \setlength{\tabcolsep}{6pt}
    \caption{Qualitative case study comparing the original \textbf{Question} with the generated \textbf{Privacy Query}. Case~1 shows that utility-critical \textbf{PII-bearing} context can be retained, while Case~2 illustrates a \textbf{failure mode} where the rewriter introduces an unentailed sensitive span.}
    \label{tab:case_study}
    \newcolumntype{L}{>{\raggedright\arraybackslash\hsize=1.2\hsize}X}
    \newcolumntype{R}{>{\raggedright\arraybackslash\hsize=0.85\hsize}X}
    \begin{tabularx}{\textwidth}{@{} L R @{}}
        \toprule
        \textbf{Question} & \textbf{Privacy Query} \\
        \midrule
        My father has \textcolor{ForestGreen}{schizophrenia}, and I'm terrified about how that might affect my relationship with my partner. On top of everything, I'm between jobs and burning through savings right now. How do I even start a conversation with my partner about my dad's condition without scaring them off?
        &
        How do I approach discussing my father's \textcolor{ForestGreen}{schizophrenia} with my partner without causing them distress? \\
        \midrule
        In a young child who suddenly became pale and stopped running, then recovered after 30 minutes and has never been cyanotic, with a normal physical examination, chest x-ray, and echocardiogram, what does an ECG pattern suggest?
        &
        In a child suddenly becoming pale and stopping while playing with a \textcolor{red}{pet Chihuahua}, no previous episodes, never cyanotic, physical and chest x-ray normal, echocardiogram normal, ECG showing the pattern on the next page, what is the diagnosis? \\
        \bottomrule
    \end{tabularx}
\end{table*}

\paragraph{Protocol.}
We conduct a fine-grained
LLM-as-a-Judge evaluation (\textsc{GPT-4o-mini}) to capture failures including semantic, inferential or combinational leakage, and subtle utility loss. Given the original
question and the rewritten query, the judge produces flags for four
privacy-leak , and returns a re-identification risk
score $r\in[0,1]$ and a utility score $u\in[0,1]$. Category definitions
and prompts are in Appendix~\ref{app:judge-rubric}.

\begin{figure}[t]
  \centering
  \includegraphics[width=1\linewidth]{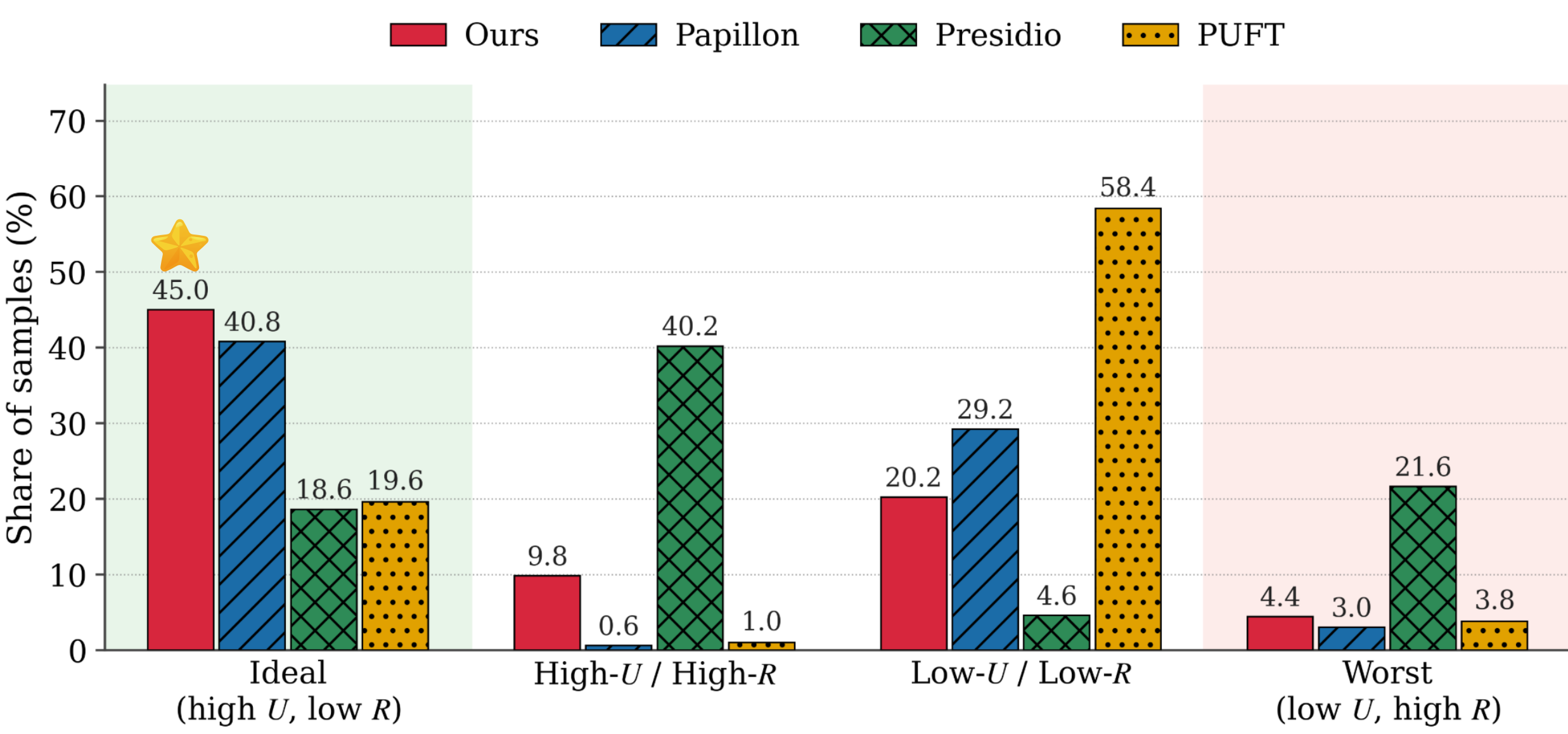}
  \caption{Joint distribution of utility score $u$ and re-identification
  risk score $r$ across the three systems.}
  \label{fig:quadrant-distribution}
\end{figure}

\paragraph{Aggregate results.}
As shown in Figure~\ref{fig:quadrant-distribution}, we observe a clear negative correlation between $r$ and $u$
(Pearson $-0.43$ overall), confirming the expected privacy--utility
tension. Within this tension, our rewrites occupy the most balanced
region of the score space: $11.2\%$ of our samples fall into the
ideal high-utility / low-risk quadrant ($u>0.7$ and $r\le 0.5$)
--- comparable to Presidio ($14.8\%$) --- while our worst-case
quadrant (low utility and high risk) contains only $3.6\%$ of samples,
less than half of Presidio's $9.0\%$. This suggests that our method
produces fewer catastrophic rewrites than Presidio while remaining more
informative than PAPILLON.

\paragraph{Privacy-failure mode analysis.}
Breaking the privacy-issue rate down by leak type
(Table~\ref{tab:leaks}) shows that the three systems fail in
qualitatively different ways. Presidio relies heavily on placeholder
substitution (e.g., \texttt{<PERSON>}, \texttt{<DATE\_TIME>}) and
consequently exhibits the highest \texttt{direct\_leak} rate
($51.8\%$ of samples), because non-name tokens such as identifiers,
diagnoses, or device names are left untouched. PAPILLON performs the
most aggressive paraphrasing and therefore reduces direct leakage to
$15.8\%$, but its rewrites retain enough correlated context to trigger
\texttt{combination\_leak} in $50.8\%$ of samples. Our method sits
between these two regimes, with the lowest severe leakage rate
(direct $+$ semantic combined) and the lowest fraction of multi-issue privacy
failures ($4.2\%$ vs.\ $7.2\%$ / $12.2\%$ / $6.9\%$).

\subsection{Qualitative Analysis}
Table~\ref{tab:case_study} illustrates two representative behaviors of our privacy rewriter. 
In Case~1, the model preserves the answer-critical clinical PII-bearing span, \emph{schizophrenia}, while removing incidental self-disclosures. Our method makes the trade-off explicit instead of over-redacting by default. In Case~2, however, the rewriter introduces an unentailed sensitive detail, ``pet Chihuahua'', despite retaining all diagnosis-critical evidence; this exposes a failure mode. A non-trivial $7.2\%$ of all rewrites introduce sensitive attributes
that are not entailed by the original question, The rate is comparable across methods
(PAPILLON $9.2\%$, Ours $6.6\%$), suggesting this is
a general hazard of LLM-driven rewriting and motivates explicit constraints against attribute hallucination in future work.

\subsection{Ablation Study}
\label{sec:exp-ablation}

\begin{table}[t]
    \centering
    \small
    \setlength{\tabcolsep}{3.2pt}
    \caption{Reward ablation with \textsc{Llama-3.1-8B-Instruct} as
    local aggregator.}
    \label{tab:reward-ablation}
    \begin{tabular}{l cc cc cc}
    \toprule
    & \multicolumn{2}{c}{Util.\ (\%)}
    & \multicolumn{2}{c}{General}
    & \multicolumn{2}{c}{Medical} \\
    \cmidrule(lr){2-3}\cmidrule(lr){4-5}\cmidrule(lr){6-7}
    Variant & Gen & Med & SLR$\downarrow$ & EA$\uparrow$ & SLR$\downarrow$ & EA$\uparrow$ \\
    \midrule
    Full reward            & \textbf{94.3} & \textbf{66.7} &  8.0 & 59.8 &  4.2 & 71.0 \\
    w/o $R_{\text{priv}}$  & 94.7 & 65.8 & 49.7 & \textbf{79.1} & 21.5 & \textbf{83.0} \\
    w/o $R_{\text{ess}}$   & 91.8 & 62.5 & \textbf{3.3} & 30.0 & \textbf{2.8} & 29.9 \\
    \bottomrule
    \end{tabular}
\end{table}

We retrain the rewriter under two reduced rewards: \textbf{w/o
$R_{\text{priv}}$} and \textbf{w/o $R_{\text{ess}}$}. As shown in
Table~\ref{tab:reward-ablation}, each reduced reward isolates one Contextual Integrity failure mode from \S\ref{sec:intro}. Without $R_{\text{priv}}$, copying the input
becomes dominant: EA peaks but SLR inflates $6$--$8\times$
(\emph{over-disclosure}). Without $R_{\text{ess}}$, near-empty
rewrites are optimal: SLR hits its floor but EA roughly halves on
General and collapses on Medical ($71.0\!\to\!29.9$)
(\emph{over-removal}). The full reward avoids both degenerate solutions. This supports the reward design as a constrained utility objective rather than a simple linear privacy--utility trade-off.

\section{Conclusion}
\label{sec:conclusion}
We present a Contextual-Integrity-based framework for
\emph{Privacy-Conscious Delegation}, where query rewriting is guided
by whether information is necessary for remote assistance. We built DelegateCI-Bench, a CI-grounded benchmark whose every instance is
annotated with explicit task-essential and task-non-essential
sensitive spans, and we turned this partition into a verifiable
reward that trains a rewriter with GRPO. The learned rewriter jointly suppresses non-essential disclosure and preserves answer-decisive context beyond prior rewriting baselines.

\section*{Limitations}
\paragraph{Operationalisation of Contextual Integrity.}
We operationalise CI as a sample-level binary partition into essential ($\mathcal{E}$) and non-essential sensitive ($\mathcal{N}$) spans, conditioned on a single recipient (a remote LLM) and a single transmission principle (task necessity). This necessarily flattens richer CI norms that are graded, recipient-, role-, and purpose-dependent, and that may evolve across multi-turn interactions. Our framework therefore does not directly model contested or ambiguous disclosures, downstream re-sharing, or norms that depend on the user's broader social context.

\paragraph{Reward design and threat model.}
Our privacy reward $R_{\text{priv}}$ flags leakage through strict lexical coverage and BERTScore-precision similarity against annotated $\mathcal{N}$ spans. It is therefore primarily sensitive to surface and near-paraphrase leakage of pre-identified sensitive content, and is weaker against (i) inferential leakage, where a non-sensitive rewrite still allows the remote LLM to infer protected attributes; (ii) cross-query linkage and longitudinal de-anonymisation across sessions.



\bibliography{custom}

\begin{thebibliography}{28}
\providecommand{\natexlab}[1]{#1}

\bibitem[{Chen et~al.(2024)Chen, Cai, Ji, Wang, Liu, Wang, Hou, and Wang}]{chen2024huatuogpt_o1}
Junying Chen, Zhenyang Cai, Ke~Ji, Xidong Wang, Wanlong Liu, Rongsheng Wang, Jianye Hou, and Benyou Wang. 2024.
\newblock \href {https://arxiv.org/abs/2412.18925} {{HuatuoGPT-o1}: Towards medical complex reasoning with {LLM}s}.
\newblock \emph{arXiv preprint arXiv:2412.18925}.

\bibitem[{Chen et~al.(2023)Chen, Li, Liu, and Yu}]{chen2023hideandseek}
Yu~Chen, Tingxin Li, Huiming Liu, and Yang Yu. 2023.
\newblock \href {https://arxiv.org/abs/2309.03057} {Hide and seek ({HaS}): A lightweight framework for prompt privacy protection}.
\newblock In \emph{arXiv preprint arXiv:2309.03057}.

\bibitem[{Cheng et~al.(2024)Cheng, Bouchacourt, and Ibrahim}]{cheng2024cibench}
Zhao Cheng, Diane Bouchacourt, and Mark Ibrahim. 2024.
\newblock \href {https://arxiv.org/abs/2409.13903} {{CI-Bench}: Benchmarking contextual integrity of {AI} assistants on synthetic data}.
\newblock \emph{arXiv preprint arXiv:2409.13903}.

\bibitem[{Dao(2024)}]{dao2023flashattention}
Tri Dao. 2024.
\newblock \href {https://arxiv.org/abs/2307.08691} {{FlashAttention-2}: Faster attention with better parallelism and work partitioning}.
\newblock In \emph{Proceedings of the International Conference on Learning Representations (ICLR)}.

\bibitem[{Hong et~al.(2024)Hong, Wang, Zhang, Li, Li, and Wang}]{hong2024dpopt}
Junyuan Hong, Jiachen~T. Wang, Chenhui Zhang, Zhangheng Li, Bo~Li, and Zhangyang Wang. 2024.
\newblock \href {https://openreview.net/forum?id=Ifz3IgsEPX} {{DP-OPT}: Make large language model your privacy-preserving prompt engineer}.
\newblock In \emph{Proceedings of the International Conference on Learning Representations (ICLR)}.

\bibitem[{Kan et~al.(2023)Kan, Qiao, Yu, Peng, Gao, and Li}]{kan2023ppts}
Zhigang Kan, Linbo Qiao, Hao Yu, Liwen Peng, Yifu Gao, and Dongsheng Li. 2023.
\newblock \href {https://arxiv.org/abs/2306.08223} {Protecting user privacy in remote conversational systems: A privacy-preserving framework based on text sanitization}.
\newblock In \emph{arXiv preprint arXiv:2306.08223}.

\bibitem[{Kwon et~al.(2023)Kwon, Li, Zhuang, Sheng, Zheng, Yu, Gonzalez, Zhang, and Stoica}]{kwon2023efficient}
Woosuk Kwon, Zhuohan Li, Siyuan Zhuang, Ying Sheng, Lianmin Zheng, Cody~Hao Yu, Joseph~E. Gonzalez, Hao Zhang, and Ion Stoica. 2023.
\newblock \href {https://doi.org/10.1145/3600006.3613165} {Efficient memory management for large language model serving with {PagedAttention}}.
\newblock In \emph{Proceedings of the 29th ACM Symposium on Operating Systems Principles (SOSP)}.

\bibitem[{Langford and Zhang(2007)}]{langford2007epoch}
John Langford and Tong Zhang. 2007.
\newblock The epoch-greedy algorithm for multi-armed bandits with side information.
\newblock In \emph{Advances in Neural Information Processing Systems (NeurIPS)}.

\bibitem[{Liu et~al.(2019)Liu, Ott, Goyal, Du, Joshi, Chen, Levy, Lewis, Zettlemoyer, and Stoyanov}]{liu2019roberta}
Yinhan Liu, Myle Ott, Naman Goyal, Jingfei Du, Mandar Joshi, Danqi Chen, Omer Levy, Mike Lewis, Luke Zettlemoyer, and Veselin Stoyanov. 2019.
\newblock \href {https://arxiv.org/abs/1907.11692} {{RoBERTa}: A robustly optimized {BERT} pretraining approach}.
\newblock \emph{arXiv preprint arXiv:1907.11692}.

\bibitem[{Lukas et~al.(2023)Lukas, Salem, Sim, Tople, Wutschitz, and Zanella-B{\'e}guelin}]{lukas2023analyzing}
Nils Lukas, Ahmed Salem, Robert Sim, Shruti Tople, Lukas Wutschitz, and Santiago Zanella-B{\'e}guelin. 2023.
\newblock \href {https://doi.org/10.1109/SP46215.2023.10179300} {Analyzing leakage of personally identifiable information in language models}.
\newblock In \emph{Proceedings of the IEEE Symposium on Security and Privacy (SP)}, pages 346--363.

\bibitem[{{Microsoft}(2021)}]{prediax}
{Microsoft}. 2021.
\newblock {Microsoft Presidio}: Context-aware, pluggable and customizable {PII} anonymization service for text and images.
\newblock \url{https://microsoft.github.io/presidio/}.

\bibitem[{Mireshghallah et~al.(2024)Mireshghallah, Kim, Zhou, Tsvetkov, Sap, Shokri, and Choi}]{mireshghallah2024can}
Niloofar Mireshghallah, Hyunwoo Kim, Xuhui Zhou, Yulia Tsvetkov, Maarten Sap, Reza Shokri, and Yejin Choi. 2024.
\newblock \href {https://openreview.net/forum?id=gmg7t8b4s0} {Can {LLM}s keep a secret? {T}esting privacy implications of language models via contextual integrity theory}.
\newblock In \emph{Proceedings of the International Conference on Learning Representations (ICLR)}.

\bibitem[{Moritz et~al.(2018)Moritz, Nishihara, Wang, Tumanov, Liaw, Liang, Elibol, Yang, Paul, Jordan, and Stoica}]{moritz2018ray}
Philipp Moritz, Robert Nishihara, Stephanie Wang, Alexey Tumanov, Richard Liaw, Eric Liang, Melih Elibol, Zongheng Yang, William Paul, Michael~I. Jordan, and Ion Stoica. 2018.
\newblock \href {https://www.usenix.org/conference/osdi18/presentation/moritz} {{Ray}: A distributed framework for emerging {AI} applications}.
\newblock In \emph{Proceedings of the 13th USENIX Symposium on Operating Systems Design and Implementation (OSDI)}, pages 561--577.

\bibitem[{Neel and Chang(2023)}]{neel2023privacy}
Seth Neel and Peter Chang. 2023.
\newblock \href {https://arxiv.org/abs/2312.06717} {Privacy issues in large language models: A survey}.
\newblock \emph{arXiv preprint arXiv:2312.06717}.

\bibitem[{Ngong et~al.(2025)Ngong, Kadhe, Wang, Murugesan, Weisz, Dhurandhar, and Ramamurthy}]{ngong2025contextual}
Ivoline Ngong, Swanand Kadhe, Hao Wang, Keerthiram Murugesan, Justin~D. Weisz, Amit Dhurandhar, and Karthikeyan~Natesan Ramamurthy. 2025.
\newblock \href {https://aclanthology.org/2025.findings-acl.1343/} {Protecting user privacy in online settings via supervised learning guided by contextual integrity}.
\newblock In \emph{Findings of the Association for Computational Linguistics: ACL 2025}. Association for Computational Linguistics.

\bibitem[{Nissenbaum(2010)}]{nissenbaum2010privacy}
Helen Nissenbaum. 2010.
\newblock \emph{Privacy in Context: Technology, Policy, and the Integrity of Social Life}.
\newblock Stanford University Press, Stanford, CA.

\bibitem[{Pil\'{a}n et~al.(2022)Pil\'{a}n, Lison, {\O}vrelid, Papadopoulou, S\'{a}nchez, and Batet}]{pilan2022text}
Ildik\'{o} Pil\'{a}n, Pierre Lison, Lilja {\O}vrelid, Anthi Papadopoulou, David S\'{a}nchez, and Montserrat Batet. 2022.
\newblock \href {https://doi.org/10.1162/coli_a_00458} {The text anonymization benchmark ({TAB}): A dedicated corpus and evaluation framework for text anonymization}.
\newblock \emph{Computational Linguistics}, 48(4):1053--1101.

\bibitem[{{Qwen Team} et~al.(2025){Qwen Team}, Yang, Yang, Zhang, Hui, Zheng, Yu, Li, Liu, Huang, Wei, Lin, Yang, Tu, Zhang, Yang, Yang, Zhou, Lin, Dang, Lu, Bao, Yang, Yu, Li, Xue, Zhang, Zhu, Men, Lin, Li, Tang, Xia, Ren, Ren, Fan, Su, Zhang, Wan, Liu, Cui, Zhang, and Qiu}]{qwen2025}
{Qwen Team}, An~Yang, Baosong Yang, Beichen Zhang, Binyuan Hui, Bo~Zheng, Bowen Yu, Chengyuan Li, Dayiheng Liu, Fei Huang, Haoran Wei, Huan Lin, Jian Yang, Jianhong Tu, Jianwei Zhang, Jianxin Yang, Jiaxi Yang, Jingren Zhou, Junyang Lin, and 24 others. 2025.
\newblock \href {https://arxiv.org/abs/2412.15115} {{Qwen2.5} technical report}.
\newblock \emph{arXiv preprint arXiv:2412.15115}.

\bibitem[{Schulman et~al.(2017)Schulman, Wolski, Dhariwal, Radford, and Klimov}]{schulman2017ppo}
John Schulman, Filip Wolski, Prafulla Dhariwal, Alec Radford, and Oleg Klimov. 2017.
\newblock \href {https://arxiv.org/abs/1707.06347} {Proximal policy optimization algorithms}.
\newblock \emph{arXiv preprint arXiv:1707.06347}.

\bibitem[{Shao et~al.(2024{\natexlab{a}})Shao, Li, Shi, Liu, and Yang}]{shao2024privacylens}
Yijia Shao, Tianshi Li, Weiyan Shi, Yanchen Liu, and Diyi Yang. 2024{\natexlab{a}}.
\newblock {PrivacyLens}: Evaluating privacy norm awareness of language models in action.
\newblock In \emph{Advances in Neural Information Processing Systems (NeurIPS)}.

\bibitem[{Shao et~al.(2024{\natexlab{b}})Shao, Wang, Zhu, Xu, Song, Zhang, Li, Wu, and Guo}]{shao2024deepseekmath}
Zhihong Shao, Peiyi Wang, Qihao Zhu, Runxin Xu, Junxiao Song, Mingchuan Zhang, Y.~K. Li, Y.~Wu, and Daya Guo. 2024{\natexlab{b}}.
\newblock \href {https://arxiv.org/abs/2402.03300} {{DeepSeekMath}: Pushing the limits of mathematical reasoning in open language models}.
\newblock \emph{arXiv preprint arXiv:2402.03300}.

\bibitem[{{ShareGPT}(2023)}]{sharegpt2023}
{ShareGPT}. 2023.
\newblock {ShareGPT}: Share your wildest {ChatGPT} conversations with one click.
\newblock \url{https://sharegpt.com/}.
\newblock Accessed: 2025-05-24.

\bibitem[{Sheng et~al.(2025)Sheng, Zhang, Ye, Wu, Zhang, Zhang, Peng, Lin, and Wu}]{sheng2024verl}
Guangming Sheng, Chi Zhang, Zilingfeng Ye, Xibin Wu, Wang Zhang, Ru~Zhang, Yanghua Peng, Haibin Lin, and Chuan Wu. 2025.
\newblock \href {https://arxiv.org/abs/2409.19256} {{HybridFlow}: A flexible and efficient {RLHF} framework}.
\newblock In \emph{Proceedings of the Twentieth European Conference on Computer Systems (EuroSys '25)}.

\bibitem[{Siyan et~al.(2025)Siyan, Raghuram, Khattab, Hirschberg, and Yu}]{papillon}
Li~Siyan, Vethavikashini~Chithrra Raghuram, Omar Khattab, Julia Hirschberg, and Zhou Yu. 2025.
\newblock \href {https://arxiv.org/abs/2410.17127} {{PAPILLON}: Privacy preservation from {I}nternet-based and {L}ocal {L}anguage {M}odel{O}rchestratio{N}}.
\newblock In \emph{Proceedings of the 2025 Conference of the Nations of the Americas Chapter of the Association for Computational Linguistics (NAACL)}.

\bibitem[{Skalse et~al.(2022)Skalse, Howe, Krasheninnikov, and Krueger}]{skalse2022reward}
Joar Skalse, Nikolaus H.~R. Howe, Dmitrii Krasheninnikov, and David Krueger. 2022.
\newblock \href {https://arxiv.org/abs/2209.13085} {Defining and characterizing reward hacking}.
\newblock In \emph{Advances in Neural Information Processing Systems (NeurIPS)}.

\bibitem[{Tang et~al.(2024)Tang, Shin, Inan, Manoel, Mireshghallah, Lin, Gopi, Kulkarni, and Sim}]{tang2024privacy}
Xinyu Tang, Richard Shin, Huseyin~A. Inan, Andre Manoel, Fatemehsadat Mireshghallah, Zinan Lin, Sivakanth Gopi, Janardhan Kulkarni, and Robert Sim. 2024.
\newblock \href {https://openreview.net/forum?id=oZtt0pRnOl} {Privacy-preserving in-context learning with differentially private few-shot generation}.
\newblock In \emph{Proceedings of the International Conference on Learning Representations (ICLR)}.

\bibitem[{Zhang et~al.(2020)Zhang, Kishore, Wu, Weinberger, and Artzi}]{zhang2020bertscore}
Tianyi Zhang, Varsha Kishore, Felix Wu, Kilian~Q. Weinberger, and Yoav Artzi. 2020.
\newblock \href {https://openreview.net/forum?id=SkeHuCVFDr} {{BERTScore}: Evaluating text generation with {BERT}}.
\newblock In \emph{Proceedings of the International Conference on Learning Representations (ICLR)}.

\bibitem[{Zhao et~al.(2024)Zhao, Ren, Hessel, Cardie, Choi, and Deng}]{zhao2024wildchat}
Wenting Zhao, Xiang Ren, Jack Hessel, Claire Cardie, Yejin Choi, and Yuntian Deng. 2024.
\newblock \href {https://openreview.net/forum?id=Bl8u7ZRlbM} {{WildChat}: 1{M} {ChatGPT} interaction logs in the wild}.
\newblock In \emph{Proceedings of the International Conference on Learning Representations (ICLR)}.

\end{thebibliography}

\appendix

\section{Setup Details}
\label{app:training}
\paragraph{Hardware and software.}
All training and evaluation are run on a single node with $4{\times}$ NVIDIA A100 80GB PCIe GPUs. Our implementation is built on the VERL framework~\citep{sheng2024verl} for reinforcement learning, with vLLM~\citep{kwon2023efficient} as the rollout engine and Ray~\citep{moritz2018ray} for distributed orchestration. The actor is sharded with PyTorch native FSDP and uses FlashAttention~2~\citep{dao2023flashattention} together with gradient checkpointing. The reward module uses \texttt{bert-score}~\citep{zhang2020bertscore} with RoBERTa-large~\citep{liu2019roberta}, layer~17, and \texttt{rescale\_with\_baseline=True}, following the official recommendation for English. Exact library versions are released with our code.

\paragraph{Model parameters.}
We fine-tune Qwen2.5-3B-Instruct~\citep{qwen2025} using Group Relative Policy Optimization (GRPO)~\citep{shao2024deepseekmath}. Training uses a global batch size of $24$ prompts, a PPO mini-batch size of $24$, and a per-GPU micro-batch size of $6$. For each prompt we sample $n{=}8$ rollouts with a generation temperature of $0.7$. The actor learning rate is $1{\times}10^{-6}$ with a $3\%$ linear warm-up and a constant schedule afterwards, weight decay $0.01$, and gradient clipping at $1.0$. We adopt the low-variance KL estimator (\texttt{low\_var\_kl}) between the actor and a frozen reference policy, with a KL loss coefficient of $0.005$, applied only to the policy loss and not to the reward. The PPO clip ratio is $0.2$ and the entropy coefficient is $0$. The context window is set to $1{,}024$ tokens for the prompt and $512$ tokens for the response. The batched reward function (\S\ref{sec:reward}) is invoked with a BERTScore batch size of $128$. We train for a maximum of $10$ epochs and select the best checkpoint by validation performance evaluated every $80$ optimizer steps.

\section{Baseline Implementation Details}
\label{app:baselines}
\paragraph{Direct LLM.}
Direct LLM directly sends the original, unmodified user query to the
remote model and uses the remote response as the final answer. This
setting does not provide privacy protection and is therefore not
included as a protected method on the privacy axis. Instead, it serves
as an upper-bound reference for remote-model utility.

\paragraph{Local Only.}
Local Only keeps the entire interaction on the trusted side by sending
the original user query directly to the frozen local aggregator, without
query rewriting or remote-model access. This setting exposes no query
content to the remote model and serves as a lower-bound reference for
utility under fully local inference.

We provide implementation details for all baselines compared in
Section~\ref{sec:exp-main}. Unless otherwise specified, all baselines
share the same remote model (\textsc{GPT-4o}) and the same evaluation
protocol to ensure a fair comparison; only the local
privacy-preserving component differs across systems. For baselines
that involve a local model (\textsc{PAPILLON} and \textsc{PUFT}), we
sweep over the same three local aggregators used in our main
experiments (\S\ref{sec:exp-setup}): \textsc{Qwen-2.5-1.5B-Instruct},
\textsc{Qwen-2.5-7B-Instruct}, and \textsc{Llama-3.1-8B-Instruct}.

\paragraph{Presidio~\citep{prediax}.}
Presidio is a rule- and NER-based PII detection and anonymization
baseline. We use the off-the-shelf Presidio Analyzer to detect
sensitive entities in the input query and the Presidio Anonymizer to
replace detected spans with anonymized placeholders before sending the
query to the remote model. The same remote model is then queried with
the anonymized prompt, and its output is used as the final response.
This baseline evaluates whether standard automatic PII redaction is
sufficient to preserve privacy while maintaining downstream response
quality.

\paragraph{PAPILLON~\citep{papillon}.}
PAPILLON is a privacy-preserving model-ensemble baseline for
privacy-conscious delegation. It is a prompting-based pipeline and
does not involve additional training or fine-tuning. Following the
original implementation, we instantiate two local-model modules: a
\emph{Prompt Creator}, which rewrites the user query into a
privacy-preserving prompt for the remote model, and an
\emph{Information Aggregator}, which combines the original user query
with the remote model's response to produce the final answer. In our
experiments, we keep the original PAPILLON prompts unchanged and only
vary the local backbone model used for both the creator and
aggregator. The remote model is fixed across runs for fair
comparison.

\paragraph{PUFT~\citep{ngong2025contextual}.}
PUFT is a prompt-based local LLM method for privacy-preserving prompt
reformulation, without any training or fine-tuning. Following the
original framework, we implement it as a local intermediary between
the user prompt and the target conversational agent. Each input is
processed through three stages: context identification, sensitive
information classification, and prompt reformulation. The model first
infers the interaction domain and user task, then separates sensitive
attributes into essential information required for task completion
and non-essential information that should be protected. When
non-essential sensitive information is detected, the model removes,
generalizes, or rewrites such content while preserving the original
intent. As with PAPILLON, we vary only the local backbone over the
three aggregators listed above and keep all prompts and the remote
model fixed.

\section{DelegateCI-Bench Construction Details}
\label{app:judge}

This appendix enumerates the sampler axes (Tables~\ref{tab:axes-domain-task}
and~\ref{tab:axes-attr}), lists the generator/judge/reformulator prompt
templates, and defines the consistency-judge acceptance rule for comparing
$(\hat{\mathcal{E}},\hat{\mathcal{N}})$ with the sampled partition
$(\mathcal{E},\mathcal{N})$.

\subsection{Coverage of the structured sampler}
\label{app:judge:coverage}

Each sample is a joint draw $(d, t, \mathcal{S}_E, \mathcal{S}_N, s, y, \tau)$ over the axes in Tables~\ref{tab:axes-domain-task} and~\ref{tab:axes-attr}; $\mathcal{S}_E$ and $\mathcal{S}_N$ come from the 30-attribute pool, and $\tau$ is generator-only.

\begin{table}[t]
  \centering
  \footnotesize
  \setlength{\tabcolsep}{4pt}
  \renewcommand{\arraystretch}{1.05}
  \caption{Domain, task and stylistic axes used by the sampler.}
  \label{tab:axes-domain-task}
  \begin{tabular}{p{0.22\linewidth} p{0.70\linewidth}}
    \toprule
    \textbf{Axis (\#)} & \textbf{Values} \\
    \midrule
    Domain (11) &
      Health\,\&\,Wellness, Financial\,\&\,Corporate,
      Employment\,\&\,Applications, Academic\,\&\,Education, Legal,
      Personal Relationships, Travel, Hobbies\,\&\,Habits,
      Sexual\,\&\,Erotic, Politics, Religion. \\
    \midrule
    Task type (20) &
      Summarisation, Prompt Generation for AI Models,
      Story\,\&\,Script Generation, Song\,\&\,Poem Generation,
      Character Description Generation, Code Generation,
      Code Editing\,\&\,Debugging, Communication Generation,
      Non-Fictional Document Generation, Text Editing,
      Comparison/Ranking/Recommendation,
      Brainstorming\,\&\,Idea Generation, Information Retrieval,
      Problem Solving, Explanation\,\&\,Practical Advice,
      Personal Advice, Back-and-forth Role Playing,
      Multiple-Choice QA, Translation, General Chitchat. \\
    \midrule
    Subject (5) &
      \emph{self}, \emph{family}, \emph{friend},
      \emph{coworker}, \emph{third party}. \\
    \midrule
    Disclosure style (4) &
      \emph{direct} (short factual statements),
      \emph{narrative} (mini-story),
      \emph{implicit} (inferable details),
      \emph{multi-sentence} (scattered details). \\
    \midrule
    Tone (6) &
      neutral, frustrated, casual, urgent, formal, anxious. \\
    \bottomrule
  \end{tabular}
\end{table}

\begin{table}[t]
  \centering
  \footnotesize
  \setlength{\tabcolsep}{4pt}
  \renewcommand{\arraystretch}{1.05}
  \caption{Sensitive-attribute taxonomy inherited from
  \citealp{ngong2025contextual}.}
  \label{tab:axes-attr}
  \begin{tabular}{p{0.22\linewidth} p{0.70\linewidth}}
    \toprule
    \textbf{Family} & \textbf{Attributes} \\
    \midrule
    Identifiers & name, email, phone\_number, address, ssn,
                  driver\_license, dates. \\
    Demographics & age, gender, ethnicity. \\
    Health      & physical\_health, mental\_health, medications,
                  allergies, disabilities, family\_history,
                  exercise\_hours, smoker, diet\_type. \\
    Beliefs/Identity & religious\_beliefs, sexual\_orientation. \\
    Lifestyle   & favorite\_food, favorite\_hobbies, pet\_ownership,
                  movie\_prefs, vacation\_prefs. \\
    Socio-economic & financial\_situation, employment, legal,
                     relationship\_status. \\
    \bottomrule
  \end{tabular}
\end{table}

\paragraph{Domain $\to$ essential pool and task essential policy.}
Two side-tables determine whether an attribute may be essential:
a domain pool $\mathcal{P}_E(d)$ and a task policy
$\pi(t)\in\{\textsc{none},\textsc{domain},\textsc{addressed-person}\}$.
\textsc{none} (Code Generation, Translation, Summarisation, MCQ,
Information Retrieval) sets $\mathcal{E}=\emptyset$; \textsc{domain}
(Personal Advice, Problem Solving) draws $|\mathcal{E}|\!\in\!\{1,2\}$
from $\mathcal{P}_E(d)$; \textsc{addressed-person} (Communication,
Story/Poem, Character Description) forces \texttt{name} into
$\mathcal{E}$ and may add one domain-pool attribute.

\paragraph{Confusion injection.}
With probability $\rho_{\text{conf}}{=}0.15$ the sampler emits a
confusion specification. \textsc{shared-attribute} ($60\%$) places the
same attribute type in both $\mathcal{E}$ and $\mathcal{N}$ with distinct
subjects, tagging the $\mathcal{N}$ side with a third-party relation.
\textsc{over-removal-trap} ($40\%$) forces the task to
\textsc{addressed-person}, making an otherwise sensitive-looking
attribute answer-bearing.

\subsection{Generator, judge and reformulator prompts}
\label{app:judge:prompts}

Generator runs use Claude Sonnet~4.6 at temperature $0.95$; the judge
uses GPT-4o at temperature $0.0$ in JSON mode; the reformulator uses
GPT-4o at temperature $0.2$.


\subsection{Generator prompt.}
\begin{promptbox}{Generator prompt.}
{\fontsize{9pt}{10pt}\selectfont

\noindent\textbf{System}\\
You are a data-synthesis assistant that crafts realistic, natural-sounding user queries for an AI chatbot. Output ONLY the user query text -- no preamble, JSON, quotes, or labels.

\noindent\textbf{User}\\
Compose a single realistic user message.

\noindent\textbf{Context}\\
domain \texttt{\{domain\}}, task \texttt{\{task\}} (or task brief \texttt{\{task\_brief\}} when no sensitive attribute is essential), subject framing \texttt{\{subject\}}, writing style \texttt{\{style\}}, tone \texttt{\{tone\}}.

\noindent\textbf{Constraints}
\begin{itemize}\setlength\itemsep{0pt}\setlength\parskip{0pt}\setlength\topsep{1pt}
\item The query MUST naturally include EVERY essential and non-essential item listed; values may be rephrased but the underlying facts must stay inferable.
\item Essential items are necessary to answer; removing them breaks the request. Non-essential items are user ``oversharing'' and must be safely removable.
\item Do NOT add new PII beyond the listed items. Length 30--150 words; sound like a real user.
\item Output ONLY the final message text.
\end{itemize}

\noindent\textbf{Confusion cues (conditional)}\\
\texttt{shared\_attr}: the same attribute type appears in both buckets; a \texttt{[SUBJECT NOTE]} marks the third-party owner, and the writing must disambiguate ownership. \\
\texttt{over\_removal\_trap}: attributes that look sensitive (e.g., a name) are essential because the task is addressed to that person; stripping them must break the request.

\noindent\textbf{Inputs appended to the prompt}\\
ESSENTIAL information: \texttt{\{e\_list\}}\\
NON-ESSENTIAL sensitive information: \texttt{\{ne\_list\}}\\
Each item is formatted as \texttt{- attr (desc): "value" [SUBJECT NOTE: ...]}.
}
\end{promptbox}

\subsection{Consistency judge prompt.}
\begin{promptbox}{Consistency judge prompt.}
{\fontsize{9pt}{10pt}\selectfont

\noindent\textbf{System}\\
You are an evaluator for question-answering systems.

\noindent\textbf{Criteria}\\
Determine whether the system-generated answer aligns with the provided gold (reference) answers. Mark as \textbf{1} if the answer (i) contains the same key information, (ii) expresses the same meaning even with different wording, and (iii) is factually consistent with the gold answers; otherwise \textbf{0}.

\noindent\textbf{Output}\\
Reply with \texttt{"1"} or \texttt{"0"} only.
}
\end{promptbox}
\subsection{Reformulator prompt.}
\begin{promptbox}{Reformulator prompt.}
{\fontsize{9pt}{10pt}\selectfont

\noindent\textbf{System}\\
You are a privacy-conscious assistant that reformulates user prompts to preserve contextual privacy while keeping the original intent and structure. Output the reformulated prompt only -- one line, no JSON, quotes, or labels.

\noindent\textbf{User}\\
Reformulate the message so that: (i) the request type is preserved (question stays a question); (ii) ALL essential information is kept, possibly generalized but still informative; (iii) non-essential information is removed or generalized away; (iv) the model rewrites only, never answers; (v) the output is fluent and natural.

\noindent\textbf{Few-shot examples}
\begin{itemize}\setlength\itemsep{0pt}\setlength\parskip{0pt}\setlength\topsep{1pt}
\item \textit{Orig:} ``What's a message that I, as a cool college student, can say to someone whom I want to meet?'' $\rightarrow$ \textit{Ref:} ``What's a good way to initiate a conversation with someone I'd like to connect with?''
\item \textit{Orig:} ``I need help writing an email to Professor Smith about missing class yesterday due to food poisoning.'' $\rightarrow$ \textit{Ref:} ``I need help writing an email about a class absence due to illness.''
\item \textit{Orig:} ``Looking for advice on managing my diabetes. I take Metformin twice daily and my sister suggested trying a keto diet.'' $\rightarrow$ \textit{Ref:} ``Looking for advice on managing diabetes and considering dietary changes.''
\end{itemize}

\noindent\textbf{Inputs appended to the prompt}\\
Original: \texttt{"""\{raw\_prompt\}"""}\\
Context (intent / domain): \texttt{\{domain\}} / \texttt{\{task\}}\\
Essential information to keep: \texttt{\{e\_list\}} (or, when no sensitive essential exists, ``Essential (the task itself, must be preserved): \texttt{\{task\_brief\}}'')\\
Information to remove or generalize: \texttt{\{ne\_list\}}\\
Provide ONLY the reformulated text.
}
\end{promptbox}

\subsection{Matching rules and acceptance thresholds}
\label{app:judge:matching}

Let $\mathcal{S}_E,\mathcal{S}_N$ be the injected attribute--value sets and
$J_E,J_N$ the judge strings returned under \texttt{essential\_attributes}
and \texttt{nonessential\_attributes}. We accept a generated sample in
three steps.

\paragraph{Step 1: value matching.}
A sampled value $v$ matches a judge string $j$ if either content-token
recall is at least $0.5$,
\[
  |T(v)\cap T(j)|/|T(v)| \ge 0.5,
\]
or, optionally, sentence-embedding cosine similarity is at least $0.65$.
Attribute types are ignored, so paraphrases and tense changes can still
match. The token-only and embedding-augmented settings agree on $97.4\%$
of a 500-sample dev split.

\paragraph{Step 2: side recall.}
We score how many sampled values the judge placed on the correct side:
\[
\begin{aligned}
\widehat r_E &= \frac{|\{v\in \mathcal{S}_E : \exists j\in J_E,\ \text{match}(v,j)\}|}{|\mathcal{S}_E|},\\[2pt]
\widehat r_N &= \frac{|\{v\in \mathcal{S}_N : \exists j\in J_N,\ \text{match}(v,j)\}|}{|\mathcal{S}_N|}.
\end{aligned}
\]
For \textsc{none}-policy samples, where $\mathcal{S}_E=\emptyset$, we set
$\widehat r_E\equiv1$. We separately log swapped assignments and residual
judge-flagged spans
$\Delta_N=J_N\setminus(\mathcal{S}_E\cup\mathcal{S}_N)$; the latter are
treated as model-leaked PII and folded into $\mathcal{N}$.

\paragraph{Step 3: acceptance.}
The final score is
\[
\phi =
\begin{cases}
\min(\widehat r_E,\widehat r_N), & \mathcal{S}_E\neq\emptyset,\\
\widehat r_N\bigl(1-0.5\cdot\text{promotion-rate}\bigr), & \mathcal{S}_E=\emptyset.
\end{cases}
\]
Samples with $\phi\ge0.8$ are accepted (\textsc{pass}); samples with
$0.5\le\phi<0.8$ are kept only for hard mining (\textsc{soft-fail}); the
rest are discarded (\textsc{hard-fail}).

\paragraph{Pipeline yield.}
With Claude Sonnet~4.6 as generator and GPT-4o as judge, the thresholds
above yield an $\approx85\%$ \textsc{pass} rate and $<\!5\%$
\textsc{hard-fail}; the reverse-synthesis split of $\mathcal{B}$ keeps
only \textsc{pass} samples.


\section{LLM-as-a-Judge Rubric}
\label{app:judge-rubric}
This appendix specifies the LLM-as-a-Judge protocol used in
\S\ref{sec:llm-judge}: judge model and decoding, the inputs given to
the judge, the leak and utility taxonomies, the scoring scales, and the
full judge prompt. The same prompt and decoding configuration are used
for all three systems compared in the main paper (\textbf{Ours},
\emph{PAPILLON} and \emph{Presidio}); only the rewritten query
\texttt{priv\_query} differs across systems.

\paragraph{Judge model and decoding.}
The judge is \textsc{GPT-4o-mini} accessed through an internal proxy.
We use \texttt{temperature}~$=0.1$, default \texttt{top\_p}, and a
single sample per item. The model is queried in JSON-object mode with
\texttt{response\_format=\{"type":"json\_object"\}}, so the judge must
return a strict JSON dictionary; any output that fails to parse is
discarded and re-queried up to three times with exponential backoff.
The same evaluation prompt is reused across all systems so that the
comparison is prompt-invariant.

\paragraph{Inputs to the judge.}
For every test instance the judge receives a JSON payload with four
fields:
\begin{itemize}
  \itemsep0pt
  \item \texttt{question}: the original user request.
  \item \texttt{priv\_query}: the rewritten query produced by the
        system under evaluation.
  \item \texttt{ea\_items}: the set of essential, answer-bearing spans
        $\mathcal{E}$ from the benchmark, used as the primary utility
        reference. We merge the dataset fields \texttt{ea},
        \texttt{origin\_ea} and \texttt{key\_spans} and de-duplicate
        case-insensitively to obtain a single list.
  \item \texttt{ne\_items}: the set of non-essential sensitive spans
        $\mathcal{N}$, used as a weak hint about what should not
        survive in \texttt{priv\_query}.
\end{itemize}
The judge is explicitly instructed to use \texttt{ea\_items} as the
primary utility reference (paraphrases, generalisations and split
mentions count as preserved), to use \texttt{ne\_items} only as hints
rather than as a checklist, and to treat content that is not entailed
by \texttt{question} as a separate hallucination signal.

\paragraph{Privacy-leak categories.}
The judge labels each privacy issue with one of four leak types,
applied at the level of individual spans of \texttt{priv\_query}:
\begin{itemize}
  \itemsep0pt
  \item \texttt{direct\_leak}: \texttt{priv\_query} contains a
        sensitive identifier or a near-verbatim copy of a sensitive
        detail from \texttt{question} (names, dates, contact strings,
        device names, identifiers).
  \item \texttt{semantic\_leak}: a sensitive detail is preserved
        through paraphrase, translation, abbreviation or
        normalisation, so that the span is no longer verbatim but
        semantically equivalent.
  \item \texttt{inferential\_leak}: the sensitive attribute is not
        explicitly written but is reasonably inferable from common
        knowledge given the retained context.
  \item \texttt{combination\_leak}: several individually innocuous
        clues jointly narrow the user, patient, organisation or
        situation to a small identifiable set; flagged only when the
        combination creates meaningful re-identification risk, not
        whenever multiple clinical or task details co-occur.
\end{itemize}
Each leak entry carries a \texttt{span} (the exact substring of
\texttt{priv\_query}, or empty for combination/inferential cases with
no single responsible span), a \texttt{source} string pointing to the
sensitive content in \texttt{question}, and a brief free-text
\texttt{reason}.

\paragraph{Utility-issue categories.}
A utility issue is raised only when an \texttt{ea\_item} (or a clearly
indispensable task need that is missing from \texttt{ea\_items}) is
\emph{materially} damaged in a way that would reduce the usefulness of
the external response. The five labels are:
\begin{itemize}
  \itemsep0pt
  \item \texttt{lost\_constraint}: an \texttt{ea\_item} or
        indispensable task need is absent from \texttt{priv\_query}.
  \item \texttt{weakened\_constraint}: an \texttt{ea\_item} is present
        but materially less specific (e.g.\ a precise dose replaced by
        ``a medication'').
  \item \texttt{over\_generalization}: \texttt{priv\_query} is so
        broad that it would likely elicit only generic background.
  \item \texttt{task\_drift}: \texttt{priv\_query} asks for a different
        kind of support than the original task requires.
  \item \texttt{added\_constraint}: \texttt{priv\_query} introduces a
        new task constraint, assumption, diagnosis, preference or
        framing that is not supported by \texttt{question}.
\end{itemize}
For multiple-choice instances, answer options are not required to be
preserved unless \texttt{ea\_items} explicitly require the option set
or decision criterion. Before flagging \texttt{lost\_constraint} or
\texttt{weakened\_constraint}, the judge is required to ask whether
the omission would actually break task-grounded support; otherwise no
issue is raised. This guards against over-flagging cosmetic rewrites.

\paragraph{Hallucinated sensitive content.}
Independently of the leak categories above, the judge lists in a
\texttt{new\_sensitive} array any span of \texttt{priv\_query} that
introduces sensitive information not entailed by \texttt{question}.
This stream is used in \S\ref{sec:llm-judge} to estimate the
attribute-hallucination rate.

\paragraph{Scoring scales.}
In addition to the categorical flags, the judge emits four scalar
fields:
\begin{itemize}
  \itemsep0pt
  \item \texttt{reid\_risk}~$\in[0,1]$: re-identification risk.
        Anchors: $0.0$ -- no meaningful identifying cues; $0.3$ --
        minor residual cues unlikely to identify the case; $0.6$ --
        clearly narrowable or linked to a small identifiable set;
        $1.0$ -- directly or almost directly identifiable.
  \item \texttt{utility\_score}~$\in[0,1]$: usefulness for the
        original task. Anchors: $1.0$ -- all \texttt{ea\_items} are
        semantically preserved; $0.7$ -- one \texttt{ea\_item} mildly
        weakened or under-specified; $0.4$ -- a major \texttt{ea\_item}
        or core knowledge need is lost; $0.0$ -- different task,
        unusable rewrite, or too generic.
  \item \texttt{verdict}~$\in\{\textsc{good},\textsc{acceptable},\textsc{bad}\}$:
        a coarse summary verdict. \textsc{good} = low privacy risk and
        high utility; \textsc{acceptable} = minor privacy or utility
        issues but still usable; \textsc{bad} = clear leak, high
        re-identification risk, hallucinated sensitive content, major
        utility loss, severe over-generalisation or task drift.
  \item \texttt{confidence}~$\in[0,1]$: the judge's self-reported
        confidence in its labels. Anchors: $0.3$ -- limited or
        ambiguous evidence; $0.6$ -- moderate evidence; $0.9$ --
        strong evidence.
\end{itemize}
All numerical analyses in \S\ref{sec:llm-judge}
(quadrant distributions, leak-rate breakdown, multi-issue rate)
operate on these fields directly; verdict labels are used only as a
coarse summary.

\paragraph{Bias controls.}
We adopt two bias controls common in LLM-as-judge work. First, every
rewrite is shown to the judge in isolation rather than alongside the
output of a competing system, so the judge cannot exhibit
\emph{position} or \emph{system-identity} bias across our three
methods. Second, the judge is never told which system produced
\texttt{priv\_query}; the JSON payload carries only
\texttt{question}, \texttt{priv\_query}, \texttt{ea\_items} and
\texttt{ne\_items}, with no system tag. To discourage
\emph{verbosity} bias, the rubric explicitly states that
preservation can take the form of paraphrases or generalised wording
and that issues should be raised only when usefulness is materially
reduced.

\paragraph{Output schema.}
The judge is required to return exactly the following JSON object,
with no surrounding prose. The parser used in our analysis pipeline
rejects any other shape.

\begin{promptbox}{Judge output schema.}
{\fontsize{8pt}{9pt}\selectfont
\begin{verbatim}
{
  "privacy_issues": [
    {
      "type": "direct_leak
             | semantic_leak
             | inferential_leak
             | combination_leak",
      "span": "substring from priv_query",
      "source": "sensitive info",
      "reason": "brief explanation"
    }
  ],
  "utility_issues": [
    {
      "type": "lost_constraint
             | weakened_constraint
             | task_drift
             | over_generalization
             | added_constraint",
      "span": "substring from priv_query",
      "source": "ea_item or task need",
      "reason": "brief explanation"
    }
  ],
  "new_sensitive": [
    {
      "span": "substring from priv_query",
      "reason": "why not entailed"
    }
  ],
  "reid_risk": 0.0,
  "utility_score": 1.0,
  "verdict": "good | acceptable | bad",
  "confidence": 0.0
}
\end{verbatim}
}
\end{promptbox}

\paragraph{Full judge prompt.}
The system prompt below is the verbatim instruction we send to
\textsc{GPT-4o-mini}. The user message is a JSON dump of
\{\texttt{question}, \texttt{priv\_query}, \texttt{ne\_items},
\texttt{ea\_items}\} together with the line ``Evaluate the following
privacy rewrite. Return ONLY the JSON object specified in the system
instructions.''

\begin{promptbox}{LLM-as-a-Judge system prompt.}
{\fontsize{9pt}{10pt}\selectfont

You are evaluating whether a privacy rewrite turns a user's original
request into a privacy-preserving query for external task support.

\medskip\noindent\textbf{Your goals:}
\begin{enumerate}
  \itemsep0pt
  \item Check whether \texttt{priv\_query} leaks sensitive information
        from \texttt{question}.
  \item Check whether \texttt{priv\_query} preserves the essential
        task attributes in \texttt{ea\_items} from a semantic
        perspective.
  \item Identify utility loss \emph{only} when an essential task
        attribute is missing, weakened, distorted, over-generalised,
        or changed in a way that would reduce the usefulness of the
        external response.
  \item Identify newly introduced sensitive information not supported
        by \texttt{question}.
\end{enumerate}

\medskip\noindent
Use \texttt{ea\_items} as the primary reference for utility. For each
\texttt{ea\_item}, decide whether it is still present in
\texttt{priv\_query} in a semantically equivalent and task-useful
form. Exact wording is not required: paraphrases, abbreviations,
generalised forms, split mentions, or implications by a more specific
retained detail can count as preserved. Use \texttt{question} only to
understand context, resolve ambiguity, and catch clearly indispensable
task needs that are missing from \texttt{ea\_items}. Do not freely
invent extra utility requirements beyond \texttt{ea\_items}.

\medskip\noindent
Use \texttt{ne\_items} as weak hints for privacy-sensitive
information. Do not mechanically match \texttt{ne\_items}, but pay
attention to whether \texttt{priv\_query} reveals sensitive
identifiers, private attributes, or identifying combinations from
\texttt{question}, especially when those details are not part of
\texttt{ea\_items}.

\medskip\noindent\textbf{Privacy issue types:}
\begin{itemize}
  \itemsep0pt
  \item \texttt{direct\_leak}: \texttt{priv\_query} directly includes a
        sensitive identifier or near-verbatim sensitive detail.
  \item \texttt{semantic\_leak}: sensitive information preserved
        through paraphrase, translation, abbreviation, normalisation,
        or equivalent wording.
  \item \texttt{inferential\_leak}: sensitive information made
        reasonably inferable from common knowledge.
  \item \texttt{combination\_leak}: multiple retained clues jointly
        narrow the user, patient, organisation or situation to a small
        identifiable set; the combination must create meaningful
        re-identification risk.
\end{itemize}

\medskip\noindent\textbf{Utility issue types:}
\begin{itemize}
  \itemsep0pt
  \item \texttt{lost\_constraint}: an \texttt{ea\_item} or
        indispensable task need is absent.
  \item \texttt{weakened\_constraint}: an \texttt{ea\_item} is present
        but materially less specific.
  \item \texttt{over\_generalization}: \texttt{priv\_query} is too
        broad and would elicit only generic background.
  \item \texttt{task\_drift}: \texttt{priv\_query} asks for a different
        kind of support than the original task requires.
  \item \texttt{added\_constraint}: \texttt{priv\_query} introduces a
        new task constraint, assumption, diagnosis, preference or
        framing not supported by \texttt{question}.
\end{itemize}

\medskip\noindent\textbf{Important rules.}
For multiple-choice questions, answer options need not be preserved
unless \texttt{ea\_items} explicitly require them. Before flagging
\texttt{lost\_constraint} or \texttt{weakened\_constraint}, ask
whether the external source would fail to provide useful task-grounded
support because of this omission; if not, do not flag it. Evidence
spans must be exact substrings from \texttt{priv\_query} when
possible. Output \emph{only} the JSON object specified above; return
no text outside the JSON.
}
\end{promptbox}

\end{document}